\documentclass[prb,twocolumn,showpacs]{revtex4-1}
\usepackage{graphicx}
\usepackage{float}
\usepackage[cp1250]{inputenc}
\usepackage{amsmath}
\usepackage{textcomp}
\usepackage{multirow}
\usepackage{float}
\begin{document}
	
	\title{Remagnetization in the array of ferromagnetic nanowires with periodic and quasiperiodic order}
	\author{K.~Szulc$^{1}$ }
	\author{F.~Lisiecki$^{2}$ }
	\author{A.~Makarov$^{3,4}$ }
	\author{M.~Zelent$^{1}$ }
	\author{P.~Ku\'{s}wik$^{2}$ }
	\author{H.~G{\l}owi\'{n}ski$^{2}$ }
	\author{J.~W.~K\l{}os$^{1,5}$ }
	\author{M.~M\"{u}nzenberg$^{5}$}
	\author{R.~Gieniusz$^{6}$ }
	\author{J.~Dubowik$^{2}$ }
	\author{F.~Stobiecki$^{2}$ }
	\author{M.~Krawczyk$^{1}$ } 
	\affiliation {
		$^{1}$Faculty of Physics, Adam Mickiewicz University in Poznan, Umultowska 85, Pozna\'{n}, 61-614, Poland \\
		$^{2}$Institute of Molecular Physics, Polish Academy of Sciences, M. Smoluchowskiego 17, Pozna\'{n}, 60-179, Poland \\
		$^{3}$School of Natural Sciences, Far Eastern Federal University, Sukhanova 8, Vladivostok, 690091, Russia \\
		$^{4}$Institute of Applied Mathematics, Far Eastern Branch, Russian Academy of Sciences, Radio 7, Vladivostok, 690041, Russia \\
		$^{5}$Institute of Physics, University of Greifswald, 17489 Greifswald, Germany\\
		$^{6}$Faculty of Physics, University of Bialystok, Ciolkowskiego 1L, Bia{\l}ystok, 15-245, Poland}
	
	\date{\today}
	
	\begin{abstract}
		We investigate experimentally and theoretically the magnetization reversal process in one-dimensional magnonic structures composed of  permalloy nanowires of the two different widths and  finite length arranged in a periodic and quasiperiodic order. The main features of the hysteresis loop are determined by different shape anisotropies of the component elements and the dipolar interactions between them. We showed, that the dipolar interactions between nanowires forming a ribbon can be controlled by change a distance between the neighboring ribbons.  The quasiperiodic order can influence the hysteresis loop by introduction additional tiny switching steps when the dipolar interactions are sufficiently strong. 
	\end{abstract}
	\pacs{}
	
	\maketitle
	
	\section{Introduction}
	
	Artificial spin systems (ASS), where large magnetic moments of the monodomain magnetic elements (MEs) significantly strength magnetostatic interactions with respect to the atomic systems, are  interesting topic of research from fundamental physics and potential application points of view.\cite{Imre205,Heyderman13,Haldar2016,Gypens2018} They allow tailoring influence of the long range interactions of the dipolar type on the ground magnetization state. One of the interesting examples is frustration of the magnetization vector orientation appearing in the array of MEs arranged in the Kagome or square lattice. There, the dipolar interactions between the magnetized MEs meeting at the vertex are modified due to the reorientation of the magnetization near the edges and proximity of the neighboring MEs.\cite{Dubowik1996, Rougemaile2013} The simple model of dipolarly coupled magnetic moments requires modification to  take properly into account strength of the coupling.\cite{Rougrmaille2011} Only recently,  few ways of the magnetostatic coupling control between the MEs in the artificial spin ice systems have been demonstrated.\cite{Perrin2016,Farhan17,Ostman18} These discoveries give additional freedom for tailoring and tuning interactions of the magnetostatic origin and to study frustrated states.
	
	 The preferential axis of the magnetization orientation in the ME made of soft magnetic materials is determined by the shape anisotropy. Its magnetization reversal is affected also by the shape of the ME ends,\cite{Kirk97,Atchison2002} rough edges,\cite{Gadbois95,DEAK200025,Bryan2004} and defects.\cite{Pfau2014} In the array of MEs, the magnetization reversal process is additionally influenced by the stray magnetic field  from all other MEs in the array.\cite{Kirk2000,Vock2017} Interestingly, in the array the magnetostatic interactions from distant elements can result in indirect coupling and even screening of the interactions from nearest elements. For this purpose, the inter-element spacing along the two perpendicular directions has been introduced in the triangular lattice of elongated MEs.\cite{Zhang2011} This large number of dependencies and competing interaction in the array makes the investigation of the remagnetization interesting.
	However, the results of experimental studies depend on the quality of the samples and defects inevitable in real samples. This makes the experimental results difficult to reproduce precisely in the numerical simulations.\cite{Kirk2001}

	Most of the investigations with ASS have been dealt with the periodic structures (PS), where every lattice point is equivalent.\cite{Heyderman13} The interesting question is, how the dipole interactions influence the magnetization reversal process in other, non-periodic types of ordered ASS, like the fractal or quasiperiodic structures (QPS). The expected hysteresis loop, a variety of magnetization reversal processes are difficult to predict.\cite{BHAT2014170,Brajuskovic2016,MONTONCELLO2017158,Shi18} Moreover, the quasiperiodic and fractal ASS offer interesting spectra of the spin wave excitations, which can be controlled and modified in magnetization reversal process, which is  potentially useful for applications in magnonics.\cite{Forestire2009,RYCHLY201818,Bhat2018,Costa2014,VALERIANO2018228,Rychly2015,Rychly2016}

	The chain of magnetic nanowires (NWs) is one of the simplest geometry, nevertheless, it allows for systematic investigation of the complexity resulting from the long-range dipole interactions in periodic and  non-periodic structures.\cite{Melo2018,Nguyen2017,Concha2018} In the paper, we investigate experimentally and theoretically the magnetization reversal process in the periodic and QPS consisting of the wide and narrow Py (Ni$_{80}$Fe$_{20}$) NWs of finite length collected into the ribbon, and also in the array of ribbons.  The separation between the ribbons, as well as a NW width,  is used to demonstrate control of the dipolar coupling between the NWs. For comparable analysis, we fabricated the reference structures in the form of periodically ordered Py NWs. We show, that the demagnetizing and the stray magnetic field distribution determine the remagnetization process in the QPS. The conclusions of our investigations can be transferred to more complex structures, like 2D quasicrystals composed of MEs.  
	
	The paper is organized as follows. In the next Section, we describe the structures under investigations, and the theoretical models used to analyze the hysteresis loops measured experimentally. In Sec.~\ref{Sec:2} we describe the results of measurements, Monte-Carlo simulations, and analysis of the stray demagnetizing field during the remagnetization in PS and QPS. In the last section, we summarize our study.

	\section{Structure and methods}
	
	\subsection{Structure}
	
	
	\begin{figure}[ht]
		\centering
		\includegraphics[width=1\linewidth]{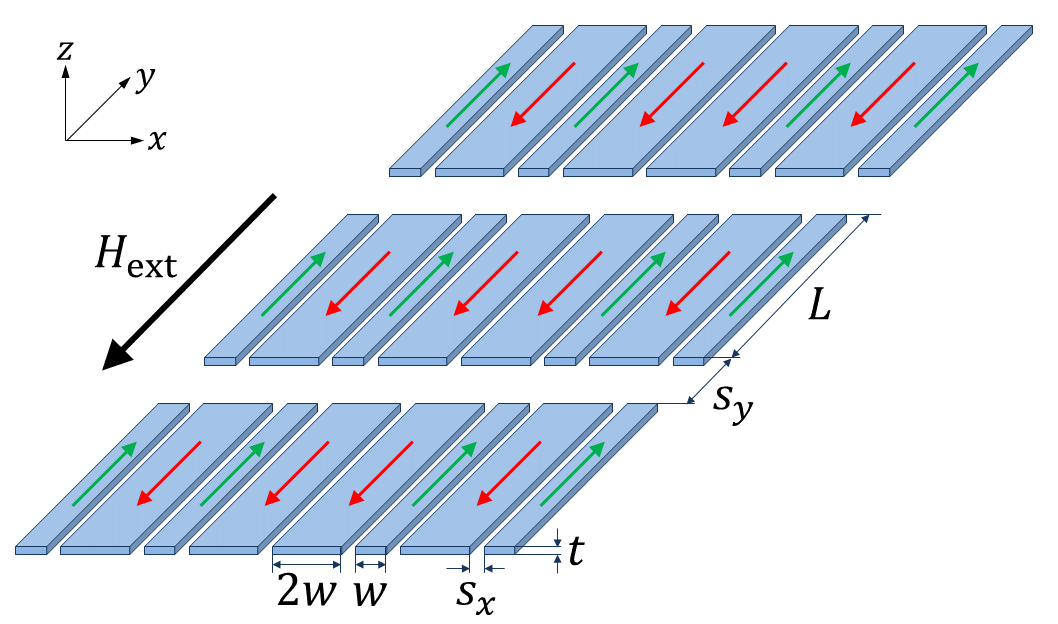}
		\caption{Section of a considered quasiperiodic structure. The flat and long magnetic NWs of thickness $t$, widths $w$ or $2w$, and length $L$ ($t\ll w\ll L$) are placed side to side and separated by air gaps of the width $s_x$. The chains of NWs form the ribbons separated from each other by the gaps of the width $s_y$. The green and red arrows show the exemplary direction of magnetization in narrow and wide NWs, respectively. The external magnetic field (black arrow) is placed along the NW axis.}
		\label{Fig:struktura}
	\end{figure}
	
	We fabricated the system of thin (thickness $t$) NWs from Py on a silicon substrate using an electron-beam lithography and a lift-off technique. The narrow ($w =$ 350 nm) and wide ($2w =$ 700 nm) NWs of finite length $L$ were arranged in the QPS according to the Fibonacci inflation rule.
	\footnote{Narrow and wide NWs can be treated as $S_0$ and $S_1$ initial Fibonacci words (initial 'sequences'), respectively. Using iteration rule $S_n = S_{n-1}+S_{n-2}$, where the symbol $+$ means the concatenation of two words (sequence), we can generate subsequent words (sequences).} The NWs are separated by $s_x =$ 100 nm wide air gaps. The total quasiperiodic sequence of NWs (100 $\mu$m long along the $x$ direction) forms the ribbon. The ribbons are repeated periodically along the $y$-axis into the array with air gaps $s_y$ between them. The structure is drawn schematically in Fig.~\ref{Fig:struktura}.
	
	In order to investigate the influence of magnetostatic interactions between the NWs on a magnetization switching, we fabricated arrays with different structural parameters. In particular, we prepared the samples with NWs of  thicknesses $t = 30$ nm and 50 nm, lengths $L = 5$ $\mu$m and 10 $\mu$m, and with the separation between the ribbons $s_y =$ 760 nm, 1.5 $\mu$m, and 10 $\mu$m. In our study, we kept the widths of NWs ($w$ and $2w$) and the separation between them ($s_x$) unchanged. The whole size of the array of ribbons was about 100 x 200 $\mu$m. 
	
\begin{figure}[ht]
		\centering
		\includegraphics[width=\linewidth]{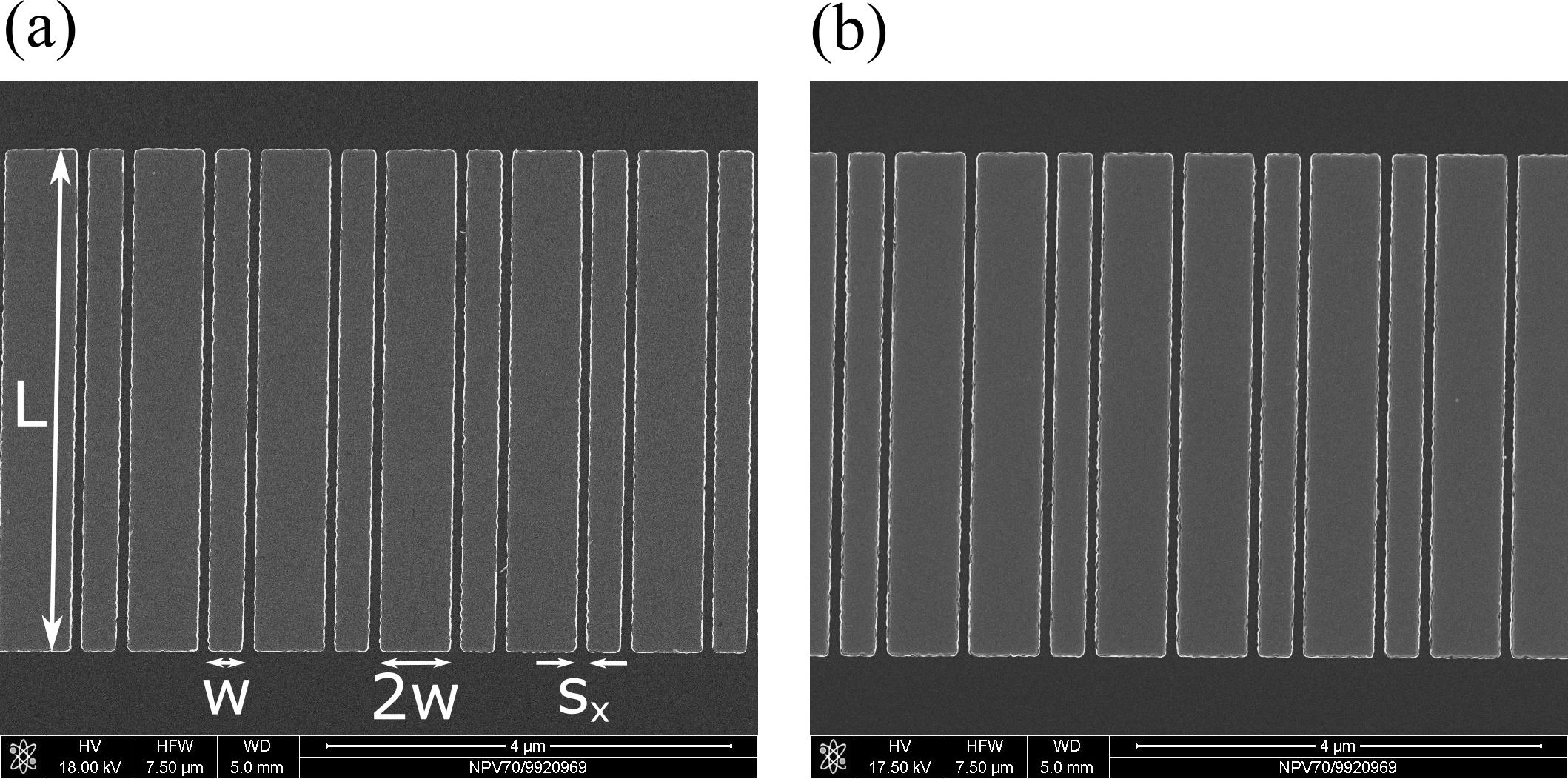}
		\caption{The scanning electron microscopy images of a single ribbon formed by (a) periodic and (b) Fibonacci sequence of NWs.}
		\label{Fig:SEM}
	\end{figure}

	We fabricated also the samples with a single ribbon and the samples with the arrays of ribbons formed by periodically ordered NWs. The PS was built by alternating repetition of the wide and narrow NWs. The fabricated structures can be seen in Fig.~\ref{Fig:SEM}(a) and (b) where the selected scanning electron microscopy (SEM) images of the Fibonacci and periodic sequences of NWs are shown, respectively.
	
	\subsection{Experimental methods}
	
	We investigated the magnetization reversal process at room temperature using the microscopy based on longitudinal magneto-optical Kerr effect (L-MOKE). The measurements were done with the aid of a wide-field polarization microscope (modified Carl Zeiss Jenapol) equipped with a CCD camera. Hysteresis loops were obtained from the evolution of magnetic domain structure recorded while the external magnetic field $H_{\text{ext}}$ is changing, being always applied along the NWs' easy axis (the $y$ axis). We collected images of the selected ribbon placed in the middle of the array during  magnetization reversal process to improve the signal-to-noise ratio. In an analysis of the magnetization switching of the individual NW in the ribbon, we took into account only wide NWs, because of the resolution of the microscope is not enough to reliably analyze the switching of the narrow NWs. 
	
	In a whole study, we observed only a full magnetization switching of the single NWs at given $H_{\text{ext}}$. At the fields $H_{\text{ext}}$ close to the switching fields, we noticed small areas, near the ends of NW, with a tilted magnetization direction (towards orientation parallel to the edge). However, this effect did not change the overall picture of abrupt magnetization switching in successive groups of the NWs with the change of $H_{\text{ext}}$.

	\subsection{Monte Carlo simulations in the Ising model}
	
	To simulate the magnetization switching in magnetic NWs arranged in PS and QPS 
	we examined the magnetic configurations and their energies using the Ising model. We considered the chain of dipolarly coupled macrospins in dependence on the strength of the external magnetic field. In our model, the wide and narrow NWs correspond to the macrospins of larger and smaller magnetic moments $m_i$, where $i$ indicates the lattice site. The external field was applied perpendicular to the chain, parallel to the macrospins. To take into account the difference in the shape anisotropy between the wide and narrow NWs, we lowered the external field at each macrospin by corresponding switching field of the NW. The switching fields $H_{\text{sw},i}$ for a single (wide or narrow) NW was extracted from micromagnetic simulations (MSs)\footnote{We used MuMax3 software to extract the $H_{\text{sw},i}$ of the isolated wide and narrow NW.}, and are listed in Table~\ref{Tab:parameters}.

	For every magnetic configuration in the considered Ising model, we compute the energy of dipolarly interacting magnetic moments in the external field:
	\begin{equation}
	E_l = k \frac{\mu_0}{4\pi} \frac{1}{2} \sum_{\substack{i,j\\ i\neq j}} \frac{m_i m_j}{\left|x_i-x_j\right|^3} - \sum_i m_i \left|H_{\text{ext}}-H_{\text{sw},i}\right|, 
	\label{Eqn:energy}
	\end{equation}
	where $m_i$ takes the values $V_i M_{\text{S}}$ and $-V_i M_{\text{S}}$ for parallel (P) and anti-parallel (AP) alignment of the magnetic moment with respect to the external magnetic field direction, respectively. The symbol $V_i$ is the volume of the $i$-th magnetic NW and $M_{\text{S}}=0.8 \times 10^6$ A/m is the saturation magnetization of Py. The symbol $x_i$ denotes the position of $i$-th magnetic moment. 
	In real samples, at the ends of NWs, a non-collinear magnetization configuration is expected which reduces the surface charges and try to close the magnetic flux inside the magnetic structure. Therefore,  the stray magnetic field generated outside the magnetic NWs will be lowered,\cite{hartmann} which can also significantly reduce the dipolar interactions between the NWs. In order to include this effect in simulations, we introduced in Eq.~(\ref{Eqn:energy}) the parameter $k$, which lowers the strength of dipolar coupling between magnetic moments in the simulating system ($0\leq k\leq 1$). 
	
	We used the Monte Carlo (MC) method based on the Metropolis algorithm \cite{metropolis,landau} to find the magnetic configurations which minimize the magnetic energy (Eq.~\ref{Eqn:energy}) when the external magnetic field is gradually changed. The details of the MC method with dipole interaction included can be found in Refs.~[\onlinecite{shevchenko1,shevchenko2}] for different models. In order to draw the hysteresis loop we changed the external magnetic field from 800 Oe to $-800$ Oe and back, with 1 Oe step. At the limiting values of $H_{\text{ext}}$ the system is magnetically saturated. In each step, i.e., for each considered value of the field, we find the magnetic configuration corresponding to the local energy minimum for all transitions from the configuration reached in the previous step.
	This quasi-adiabatic change of the magnetic configurations induced by almost continuous variation of the $H_{\text{ext}}$ allows us to determine the dependence of net magnetic moment on the external field: $M(H_{\text{ext}})/M_S = \sum_i m_i/\sum_i |m_i|$.\cite{kapitan,nefedev} The numerical calculations were performed for PS and QPS composed of 144 magnetic moments, which is the number of NWs in the experimental structures.

	\subsection{Demagnetizing field of the single nanowire}
	
	The basic element of the analytical model used in the paper is a rectangular prism of width $w$, length $L$, and thickness $t$ (Fig.~\ref{Fig:struktura}). We assume, that the NW is made of ferromagnetic material and it is homogeneously magnetized along the $y$ axis. 
	Using the Maxwell equations in magnetostatic approximation:
	\begin{eqnarray}
	\nabla\cdot (\textbf{H}_{\text{demag}} + \textbf{M}) = 0,\\
	\nabla\times \textbf{H}_{\text{demag}} = 0,
	\end{eqnarray}
	we can introduce magnetostatic potential $\textbf{H}_{\text{demag}}(\textbf{r})~=-\nabla \varphi(\textbf{r})$ and derive the general formula for $\varphi$:
	\begin{equation}
	\varphi (\textbf{r}) = -\int\limits_{V} \mathrm{d}{v'} \frac{\nabla'\cdot \textbf{M}(\textbf{r}')}{|\textbf{r}-\textbf{r}'|} + \oint\limits_{S} \mathrm{d}\sigma' \frac{\textbf{n}\cdot\textbf{M}(\textbf{r}')}{|\textbf{r}-\textbf{r}'|}, \label{Eq:Psi}
	\end{equation}
	where $V$ is volume of the NW, $S$ is a surface of the NW, and $\textbf{n}$ is the vector normal to the NW surface pointing outside.  Magnetic charges on the NW's sides perpendicular to the $y$ axis can be considered as a source of the demagnetizing and stray fields, inside and outside of the NW, respectively. In Eq. (\ref{Eq:Psi}) the part with the volume integral is equal to zero and only surface term contributes to the demagnetizing field. Eventually, formula for the field component parallel to the magnetization is described as follows:\cite{schlomann,fukushima}
	\begin{multline}
	H_{\text{demag}}^{i}(\textbf{r},\textbf{r}_{i}) = M_{\text{S}}  \sum_{\alpha,\beta,\gamma=1}^{2} (-1)^{\alpha+\beta+\gamma} \\ \times \arctan \left[ \frac{(x-x_{i}-x_\alpha)(z-z_{i}-z_\gamma)}{(y-y_{i}-y_\beta)|\textbf{r}-(\textbf{r}_{i}+\textbf{r}'_{\alpha, \beta, \gamma})|} \right],\label{Eqn:Hdemag}
	\end{multline}
	where $\textbf{r}_{i}~=~(x_{i},y_{i},z_{i})$ denotes position of the $i$-th NW, $\textbf{r}~=~(x,y,z)$, $\textbf{r}'_{\alpha,\beta,\gamma}~=~(x_{\alpha},y_{\beta},z_{\gamma})$,
	$x_1=y_1=z_1=0$, $x_2=w$, $y_2=L$, and $z_2=t$.
	We neglect the components of the demagnetizing field perpendicular to the magnetization, because their average values are equal to zero inside the stripe.
	Equation~(\ref{Eqn:Hdemag}) allows also to calculate the stray field produced by the $i$-th NW, just by taking the location $\textbf{r}$ outside of the NW.
	

	\section{Results}\label{Sec:2}
	
	\subsection{Magnetization reversal process in L-MOKE measurements}
	
\begin{figure}[ht]
		\centering
		\includegraphics[width=1\linewidth]{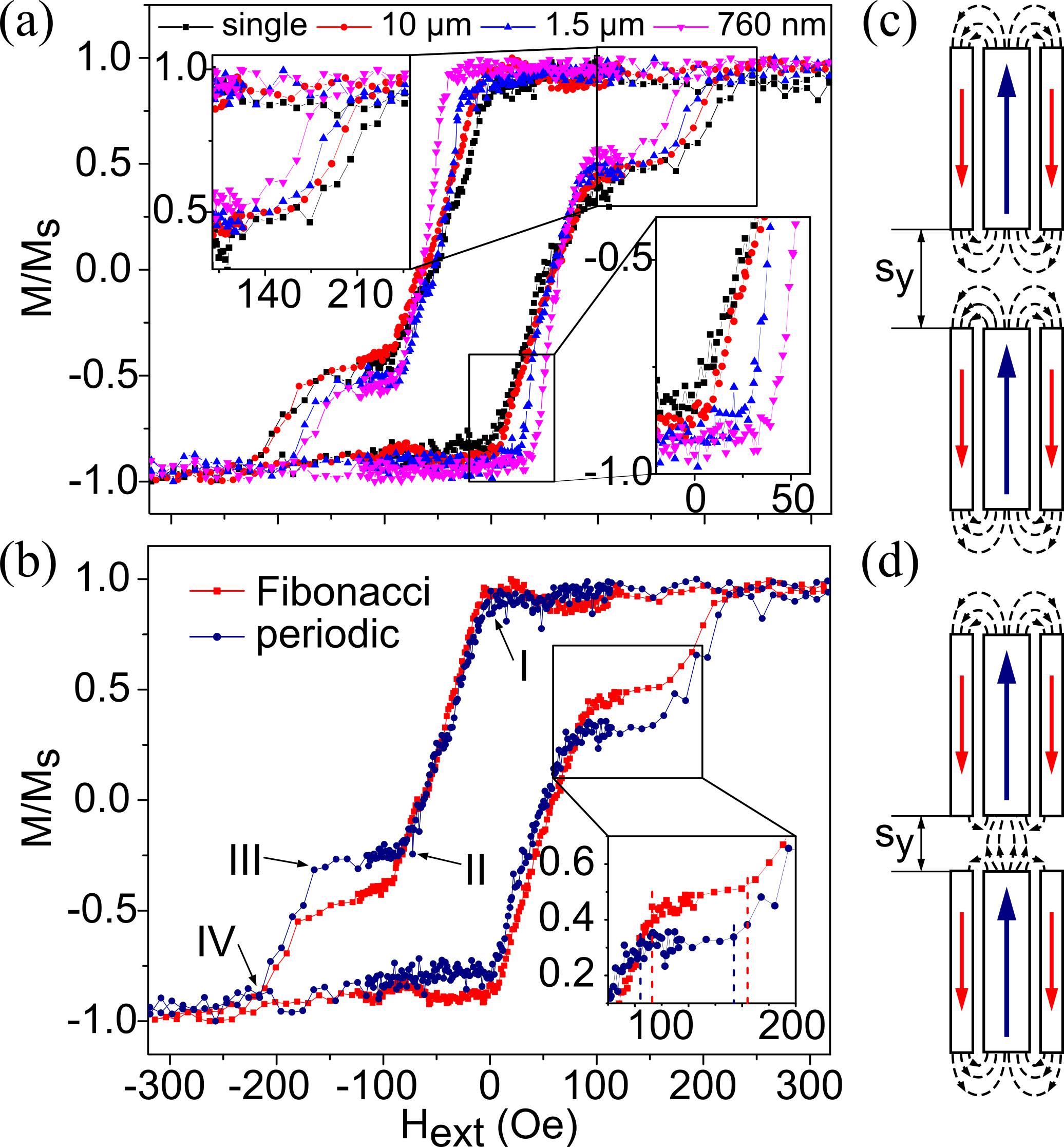}
		\caption{Comparison of the hysteresis loops measured with L-MOKE in terms of (a) the separation between ribbons for the QPS with the NWs of 5 $\mu$m length and 50 nm thickness, and (b) for Fibonacci and PS array of NWs of 5 $\mu$m length, 50 nm thickness and 10 $\mu$m separation between the ribbons. Vertical dashed lines in the inset mark the beginning and the end of the plateau. The labels I and II (III and IV) are related to switching of the wide (narrow) NWs. Scheme of the magnetic field lines from the wide NWs in the array when ribbons are well separated and close to each other are shown in (c) and (d), respectively. }
		\label{Fig:Hysteresis}
	\end{figure}

	In Fig.~\ref{Fig:Hysteresis}(a) we show comparison between the hysteresis loops for the arrays of ribbons of the Fibonacci QPS composed of NWs with 5 $\mu$m length, 50 nm thickness and for various separations between the ribbons $s_y=$ 760 nm, 1.5 $\mu$m, 10 $\mu$m. For reference, we also place in Fig.~\ref{Fig:Hysteresis}(a) the outcomes for a single ribbon. 
	
	In all hysteresis loops, the two main steps of switching are clearly observed. The lower (higher) switching field is attributed to the magnetization switching in the wide (narrow) NWs, accordingly with their smaller (larger) shape anisotropy.\cite{goolaup} In arrays consisting of dipolarly coupled NWs, the magnetostatic interaction favors the AP magnetization orientation of the neighboring NWs.\cite{gubbiotti} For such configuration the lines of the stray magnetic field  between the adjacent NWs are closed, which minimizes the magnetostatic energy of the system.\cite{bennett} Thus, the AP configuration stabilizes the system creating a plateau in the $M(H)$ dependence and the whole reversal process occurs in a wider magnetic field range. At higher switching field, we observe the transition between AP and P configuration related to the switching of narrower NWs.
	
	By decreasing the distance between the ribbons, we can move the switching field of wide stripes to higher values [see the inset in the right-bottom corner in Fig.~\ref{Fig:Hysteresis}(a)], simultaneously the fields at which the narrow NWs switch are moving to lower fields [see the inset in the left-top corner in Fig.~\ref{Fig:Hysteresis}(a)], with differences reaching several dozens Oe. 
	When the ribbons are getting closer to each other the field lines from the NWs start to link the NWs from neighboring ribbons. As a result, the magnetic flux between the NWs in the same ribbon is lowered, as it is shown schematically in Fig.~\ref{Fig:Hysteresis}(c,d). This decreases the interactions between the NWs in the ribbon and reduces the width of the plateau related to the AP magnetization configuration. Interestingly, the distance 10 $\mu$m is usually considered to be sufficient for neglecting the magnetostatic interactions.\cite{iglesias} However, we still are able to observe some noticeable differences in the switching fields between a single ribbon and an array of ribbons with air gaps separating the ribbons up to 10 $\mu$m.
	
	Furthermore, we also find that for the sequences of thinner (or longer) NWs, we observe similar changes in the switching fields as described above, resulting from the decrease of the strength of the magnetostatic interactions between NWs inside the same ribbon.\cite{fraerman}
	
	In Fig.~\ref{Fig:Hysteresis}(b) we present comparison between hysteresis loops measured for the ribbons with $s_y$ = 10 $\mu$m separation composed of 5 $\mu$m long and 50 nm thick NWs with the periodic and quasiperiodic order. Slight differences between both curves are visible. The most significant difference is a magnetization value at the plateau. This is the effect of a different ratio of the wide to narrow NWs' numbers, which takes a value 1 for PS and $(1+\sqrt{5})/2\approx$1.618 for QPS, and they correspond to plateau levels at the value 0.33 and 0.53 $M/M_{\text{S}}$, respectively. 
	The other differences, which we will later relate to the different stray magnetic fields from the different arrangements, are seen at the beginning and at the end of the plateau phase [see, the inset in Fig.~\ref{Fig:Hysteresis}(b)].
	For the quasiperiodic arrangement, we need a few Oe higher field to finish the switching of the magnetization in the wide NWs, than in a periodic structure. Also, the beginning of switching of the narrow NWs is at higher fields for QPS than for the PS.
	\begin{figure}[h]
		\centering
		\includegraphics[width=1\linewidth]{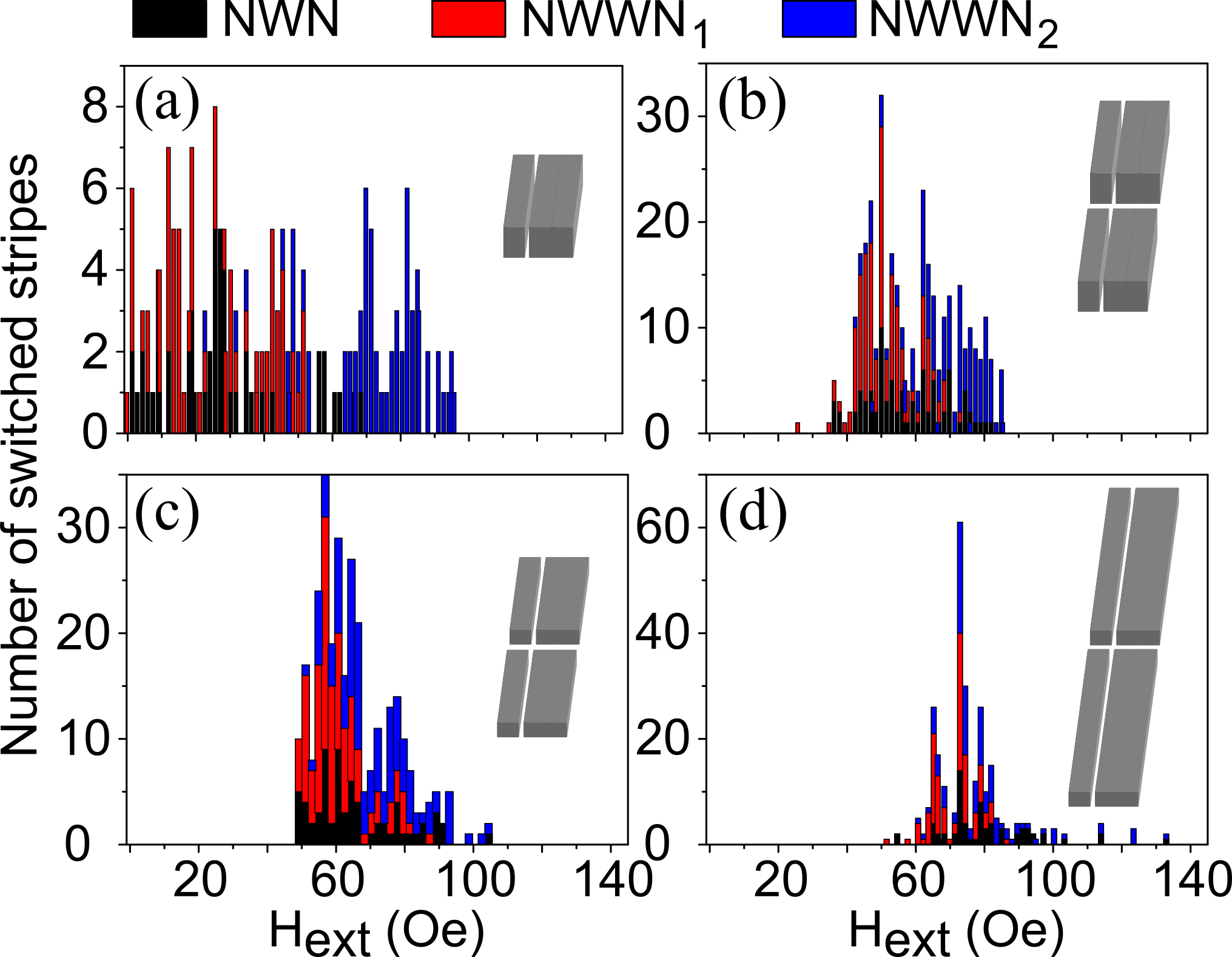}
		\caption{Number of wide NWs (of 700 nm width) which are switched in successive intervals of external field for the Fibonacci QPS differing in the NW length / NW thickness / separation between ribbons: (a) 5 $\mu$m/50 nm/$\infty$ (single ribbon), (b) 5 $\mu$m/50 nm/760 nm, (c) 5 $\mu$m/30 nm/760 nm, and (d) 10 $\mu$m/30 nm/760 nm. NWN is a wide NW between two narrow NWs, and NWWN$_1$ and NWWN$_2$ is a wide NW from a pair of two wide ones between the narrow NWs which is switched in a pair at a lower or higher field, respectively (see also the legend at the top of the Figure for color bar definition).}
		\label{Fig:Histograms}
	\end{figure}
	
	Results presented in Fig.~\ref{Fig:Histograms} show another interesting feature of the magnetization reversal process characteristic for the QPS. Let's consider the QPS, where the small external field is applied along the NWs and all NWs are initially magnetized opposite to the direction of the external field. We check, using the Kerr microscopy, how the wide NWs switch with the increase of external field (see Fig.~\ref{Fig:Histograms}). In QPS, the wide NWs can appear in pairs surrounded by narrow NWs (marked as NWWN, which stands for the consecutive wires order: narrow, wide, wide, narrow) or be left as a single wide NW surrounded by narrow NWs on both sides (NWN -- narrow, wide, narrow). Generally, the magnetization of the first NW of the pair (NWWN$_1$) switches in a similar range of the external magnetic field to a single wide NWs (NWN). However, the second NW of the pair (NWWN$_2$) switches magnetization at a higher external magnetic field than the NWN. This indicates, that there is a magnetostatic interaction between the pair of the wide NWs which introduces preferential AP configuration between wide NWs in the NWWNs. This effect is the most visible for the system with a single ribbon composed of 5 $\mu$m long and 50 nm thick NWs (Fig.~\ref{Fig:Histograms}(a)). For this structure, the magnetostatic interactions between the NWs inside the ribbon is expected to be strongest among the systems presented in Fig.~\ref{Fig:Histograms}, thus the observed effect shall diminish for other structures, where interactions are decreasing. 
	
	Indeed, in Fig.~\ref{Fig:Histograms} [and also in the inset at the right-bottom corner of Fig.~\ref{Fig:Hysteresis}(a)] it is clearly seen, that the distribution of wide NWs switching fields narrows with decreasing the distance between the ribbons, i.e., when the magnetostatic interactions between NWs inside the ribbon decreases.
	In Fig.~\ref{Fig:Histograms}(a) the second NW from the pair (NWWN$_2$) reverses mainly in the range between 60 and 100 Oe, but this range becomes narrower and overlaps with the switching fields of the rest of the wide NWs with decreasing interactions between NWs, like in Fig.~\ref{Fig:Histograms}(b). With further lowering the interactions strength between the NWs, by decreasing thickness (Fig.~\ref{Fig:Histograms}(c)) or increasing length of the NWs (Fig.~\ref{Fig:Histograms}(d)), the switching range of wide NWs narrows and differences in switching fields between NWWN$_1$ and NWWN$_2$ vanishes. These effects do not occur in a PS, where there is no two wide NWs, next to each other.
	
	We showed, that we can modify the strength of the magnetostatic interactions between the NWs in the ribbon in two ways: by changing the dimensions of the NWs (length or thickness) or by changing the distance between the ribbons. The decrease of the interactions between the NWs in the ribbon shall also decrease the differences between QPS and PS. These differences will be further investigated in the following parts of the paper with the use of the numerical MC simulations and analysis of the stray magnetic field in the systems. 
	
	\begin{table}[t]
		\caption{The switching field $H_{\text{sw}}$ obtained from micromagnetic simulations (MS); the scaling factor $\kappa$  of the magnetostatic interactions between NWs and anisotropy field $H_{\text{ani}}$ derived from the linear regression (LR) analysis using Eq.~(\ref{Eqn:linreg}) at four selected points of the hysteresis loop (see Fig.~\ref{Fig:Hysteresis}): I and II (III and IV) related to  switching of the wide (the narrow) NWs.}
		\begin{tabular}{|c|c|c|cccc|}
			\hline
			$t$ (nm)&Parameter&Method&I&II&III&IV\\ \hline            \multirow{3}{*}{30}&$\kappa$&LR&0.037&0.050&0.123&0.125\\ \cline{2-7}
			&\multirow{2}{*}{$H_{\text{sw}}$ (Oe)} &LR&51&105&104&160\\ \cline{3-7}
			&&MS&\multicolumn{2}{c}{135}&\multicolumn{2}{c|}{295}\\ \hline
			\multirow{3}{*}{50}&$\kappa$&LR&0.059&0.046&0.081&0.049\\ \cline{2-7}
			&\multirow{2}{*}{$H_{\text{sw}}$ (Oe)} &LR&42&82&123&182\\ \cline{3-7}
			&&MS&\multicolumn{2}{c}{135}&\multicolumn{2}{c|}{265}\\
			\hline
		\end{tabular}
		\label{Tab:parameters}
	\end{table}

	\subsection{Monte Carlo simulations of the remagnetization}
	
	\begin{figure}[h]
		\centering
		\includegraphics[width=1\linewidth]{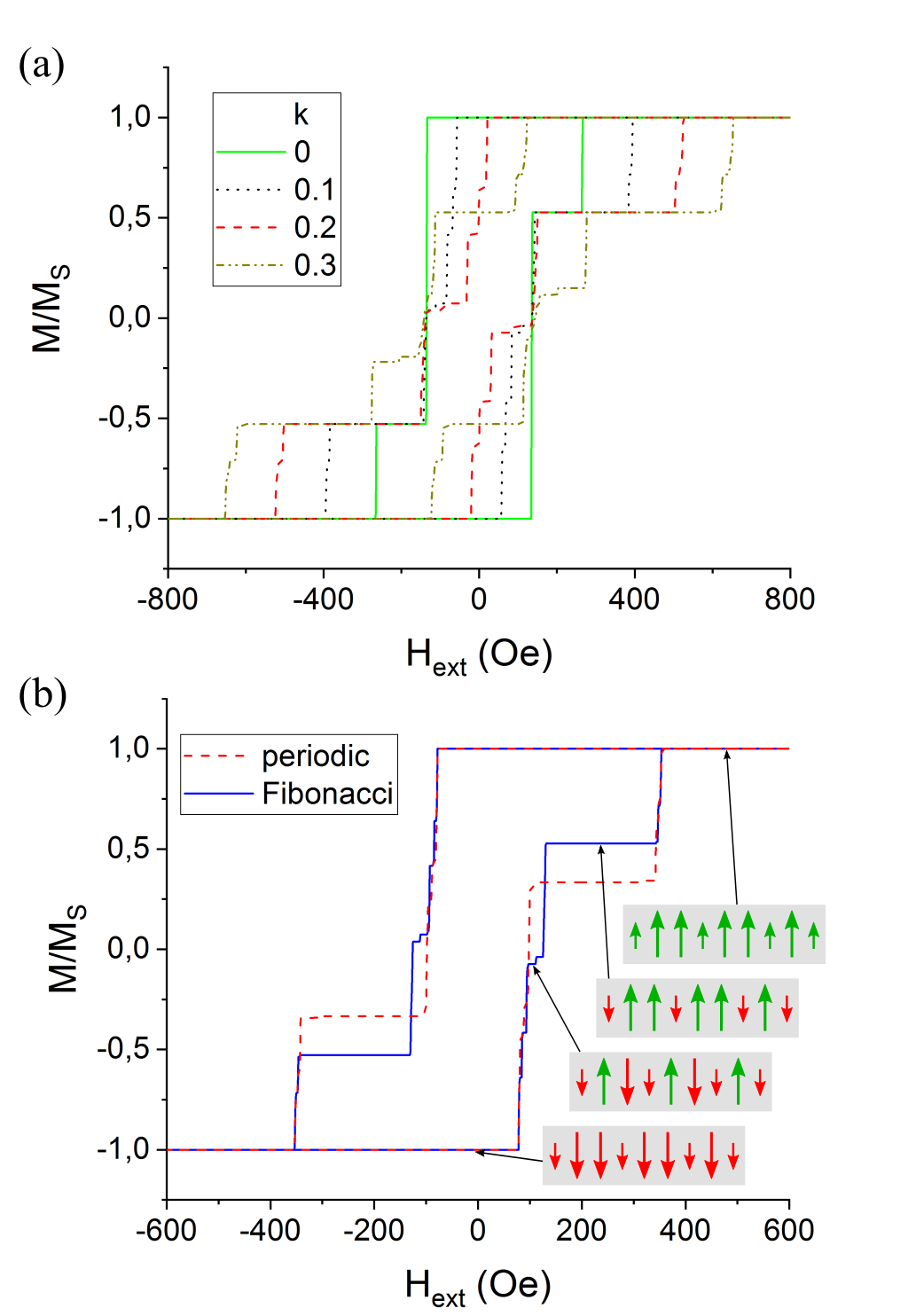}
		\caption{Comparison of the hysteresis loops obtained from MC simulations. (a) For the single Fibonacci chain of magnetic moments corresponding to 5 $\mu$m long and 50 nm thick NWs for different values of the $k$ parameter. (b) For the periodic and Fibonacci sequence of the magnetic moments, corresponding to 5 $\mu$m long and 50 nm thick NWs at $k$ fixed to 0.05.}
		\label{Fig:MC}
	\end{figure}
	
	In Fig.~\ref{Fig:MC}(a) we present the influence of dipolar interactions strength on the hysteresis loop obtained in MC simulations. For non-interacting NWs ($k=0$) the remagnetization follows the two steps process, separated by the plateau of the width equal to the difference between the switching fields of the isolated wide and narrow NWs. An increase of the parameter $k$ makes the plateau wider. Magnetization reversal process of the narrow stripes moves to higher values of $H_{\text{ext}}$ with increasing $k$, while for the wide NWs the field of the switch beginning moves to lower values. Interestingly, the beginning of the plateau (the end of the wide NWs switching) remains almost on the same position. Nevertheless, the plateau is enlarged with increasing interactions between magnetic moments. According to the MC simulations, the magnetization of the narrow NWs in PS and QPS switches at significantly higher values in reference to the values reported in the experiment (Fig.~\ref{Fig:Hysteresis}(b)). We associated this difference with too high magnetic switching fields assumed for single NW, which can be related to the regular rectangular shape used in MS, and a lack of defects. 
	
	In Fig.~\ref{Fig:MC}(a) we can  see additional narrow plateau for QPS at the level of $M=0.05 M_S$ for $k = 0.1$ and 0.2, which enlargers with $k$, and which represents magnetization state, where only the second from the pair of wide NWs (NWWN$_2$) has not yet been switched. In Fig.~\ref{Fig:Histograms} we have seen that the behavior of the experimental system is similar. Lack of a clear plateau in the experiment can be associated with defects and deviation from the rectangular shape of the NWs, which facilitate nucleation of of the reversal process and influence the switching process. Detailed inspection of Fig.~\ref{Fig:MC}(a) for $k = 0.1$ and $k= 0.2$ allows to identify also some additional steps in the reversal of the wide NWs (at $M = 0.4 M_S$ and $0.6 M_S$) and narrow NWs (at $M = 0.65 M_S$). They point at the parts of the hysteresis loop where an influence of the long-range order can be expected, whenever the effective magnetostatic interactions between NWs will be increased. Interestingly, for very strong dipolar interactions $k> 0.25$ the scenario of the remagnetization changes, see the curve for $k=0.3$ in Fig.~\ref{Fig:MC}(a). In this case, the structure tends to start the magnetization reversal process from the narrow NWs. However, such strong dipolar interactions are not accessible in our experiments. The qualitative agreement with the experimental data appeared to be for $k\approx 0.05$. 
	
	As was already discussed in the previous subsection, a decrease of the separation between ribbons results in weakening of interactions between NWs in the ribbon [see Fig.~\ref{Fig:Hysteresis}(c,d)]. The same effects should be observed with lowering the value of $k$ in MC simulations. Indeed, MC simulations confirm this hypothesis. 
	
	Comparison of hysteresis loops for the PS and QPS obtained from MC simulations for $k = 0.05$ is shown in Fig.~\ref{Fig:MC}(b). There is no additional plateau phase in the periodic structure, which is associated with lack of the pairs of wide NWs. Generally, wide NW end their remagnetization at higher $H_{\text{ext}}$ in QPS than in the PS, just as it was found in the experimental results shown in Fig.~\ref{Fig:Hysteresis}(b).

	\subsection{The structure field in the periodic and quasiperiodic sequence of nanowires}
	
	We are going now to investigate quantitatively, based on the analytical approach, the strength of dipolar interactions between selected NW and the other NWs in the structure at different points of the hysteresis loop (see Roman numerals in Fig.~\ref{Fig:Hysteresis}). This study shall give us additional information about the magnetization switching and the influence of geometrical parameters on this process.\cite{Pfau2014,Vock2017}
	
	We consider an array of rectangular prisms (Fig.~\ref{Fig:struktura}), with dimensions and separating distances being the same as in the experimental samples. We take for calculations the ribbon made of 154 NWs for PS and 144 NWs for QPS. We take three different separations between the ribbons: 760 nm, 1.5 $\mu$m, and 10 $\mu$m, which are related to a different number of ribbons in the structure: 35, 31 and 13, respectively.  
	There are two types of magnetization configurations that will be tested. First one, it is a P configuration, representing system at the beginning and at the end of magnetization reversal process--labeled as I and IV, respectively. Second, is an AP configuration, corresponding to the plateau phase obtained in experimental results, at points labeled as II and III.

	\begin{figure}[!ht]
		\includegraphics[width=1\linewidth]{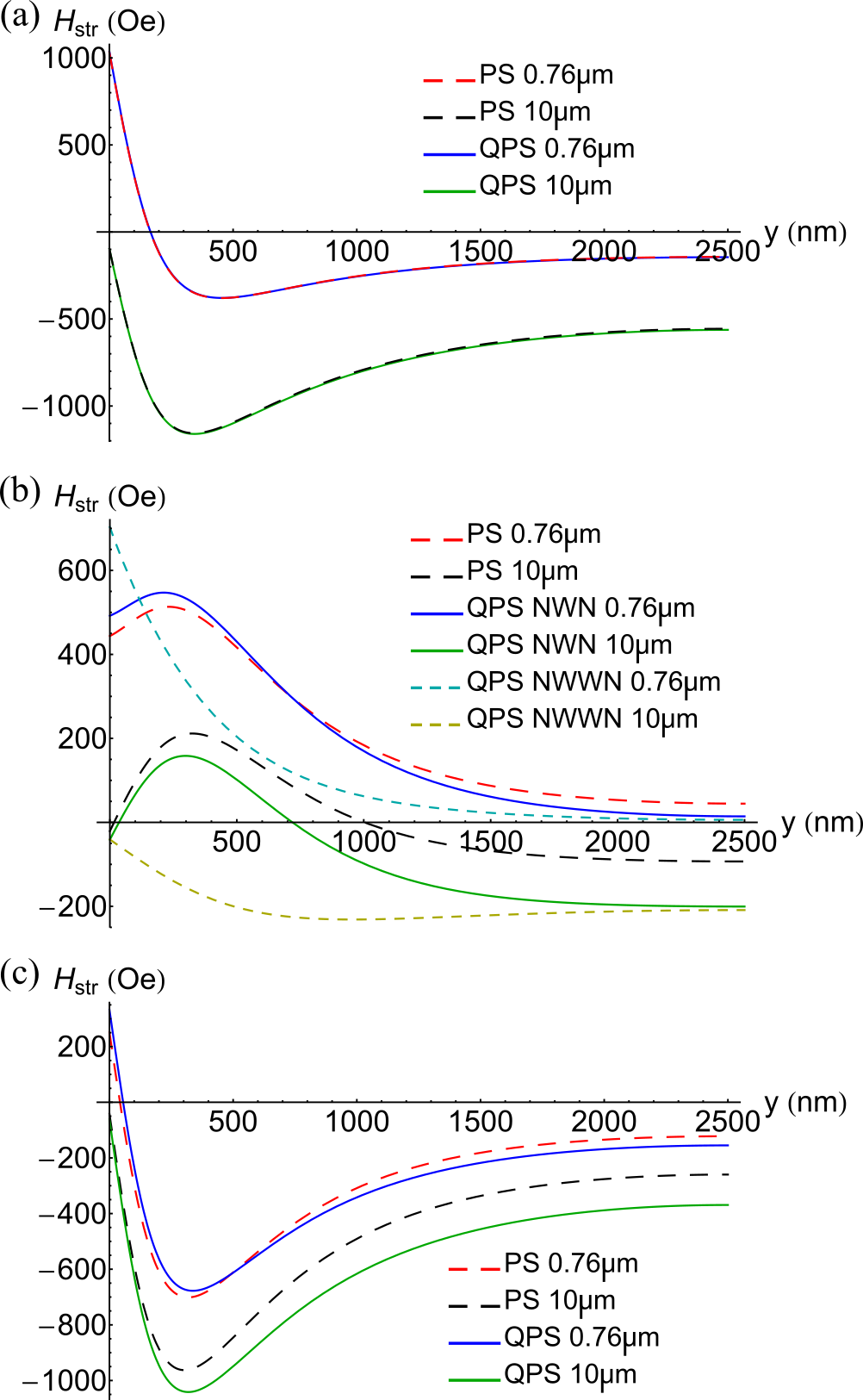}\\
		\caption{The structure field $H_{\text{str}}$ in the PS and QPS calculated for: (a) a wide stripe (at NWN and NWWN positions) with P configuration of NWs, (b) a wide stripe (at NWN and NWWN positions) with AP configuration of NWs, and (c) a narrow stripe with AP configuration of NWs. The fields in (a), (b) and (c) are related to the points I, II and III in the hysteresis loop marked in Fig. \ref{Fig:Hysteresis}(a), respectively. The results for arrays of ribbons with the separation of 10 $\mu$m and 0.76 $\mu$m are marked with different colors. We plot $H_{\text{str}}$ along the NW axis (along $y$ axis) in the middle of the NW.}
		\label{Fig:structure_field}
	\end{figure}
	
	The total magnetostatic field $H_{\text{magn}}$ can be expressed as a sum of the demagnetizing fields originating from individual NWs in the structure:
	\begin{equation}
	H_{\text{magn}} = \sum_{j}^{\text{all NWs}} H_{\text{demag}}^{j}.\label{Eq:demag_field}
	\end{equation}
	The parameter, which we select for further analysis, is the stray magnetic field  from all other NWs besides the considered $i$-th NW, and it will be called the structure field:
	\begin{equation}
	H_{\text{str}}^{i} = H_{\text{magn}} - H_{\text{demag}}^{i}.
	\label{Eq:str_field}
	\end{equation}
	This field gives the information, how much of the magnetic field inside the selected NW is present due to interaction with other elements. We remind, the Eqs.~(\ref{Eq:demag_field} and \ref{Eq:str_field}) are derived under assumption of collinear magnetic configuration inside each NW. 
	
	First, we investigate $H_{\text{str}}$ in the PS and QPS with different separations between the ribbons and for different configurations of the magnetization. We start from the P configuration, points I and IV, Fig.~\ref{Fig:structure_field}(a). There is no visible influence of the NWs order (periodic or quasiperiodic), which is according with the experimental results. Nevertheless, the increase of separation between the ribbons from 0.76 $\mu$m to 10 $\mu$m increases $H_{\text{str}}$ by at least 500 Oe. Increase of $H_{\text{str}}$ leads to decrease in $H_{\text{ext}}$ at which switching happens. This explains the changes in the switching field at the beginning and at the end of the magnetization reversal process observed experimentally in Fig. \ref{Fig:Hysteresis}(a).

	The AP configuration is represented on hysteresis loops by the plateau phase, and so it starts at the end of magnetization reversal in the wide NWs (point II in Fig.~\ref{Fig:Hysteresis}(b)), and ends at the beginning of switching narrow NWs (point III). From the profile of the structure field calculated in the wide (Fig.~\ref{Fig:structure_field}(b)) and narrow NW (Fig.~\ref{Fig:structure_field}(c)) we see that the structure field in QPS is lower than in PS for both, wide and narrow NWs, which prefers remagnetization in QPS at higher $H_{\text{ext}}$. 
	The differences in $H_{\text{str}}$ between PS and QPS are especially visible in $H_{\text{str}}$ of NWWN. The reason for this effect lies in the nearest neighbors of the wide NW in the QPS. 
	In the QPS, the second stripe from the pair of wide NWs (NWWN$_2$) has lower $H_{\text{str}}$, than the single wide stripe (NWN). This difference of $H_{\text{str}}$ gives rise to additional narrow plateau for QPS at the level of $M=0.05 M_S$.

	\subsection{Switching fields---theory and experiment}
	
	The total magnetic field inside selected NW is a sum of the external, magnetostatic and shape anisotropy field: 
	\begin{equation}
	H_{\text{tot}} = H_{\text{ext}} + H_{\text{magn}} - H_{\text{ani}}.
	\label{Eqn:totalfield0}
	\end{equation}
	We introduce the switching field in the way as it was computed in MS. It can be described as a function of internal magnetic fields:
	\begin{equation}
	H_{\text{sw}} = H_{\text{ani}} - H_{\text{demag}}\label{Eq:sw}.
	\end{equation}
	Using Eq.~(\ref{Eq:sw}) and Eq.~(\ref{Eq:str_field}) we can rewrite Eq.~(\ref{Eqn:totalfield0}) to the following form:
	\begin{equation}
	H_{\text{tot}} = H_{\text{ext}} + H_{\text{str}} - H_{\text{sw}}.
	\label{Eqn:totalfield}
	\end{equation}
	The experimental values of the external magnetic field $H_{\text{ext}}$ at which selected NW switches and the structure field $H_{\text{str}}$ obtained from the analytical model can be related to each other by the following equation:
	\begin{equation}
	H_{\text{ext}} (H_{\text{str}}^{\text{av}}) = -\kappa H_{\text{str}}^{\text{av}} + H_{\text{sw}}.
	\label{Eqn:linreg}
	\end{equation}
	In Eq.~(\ref{Eqn:linreg}) we used assumption that $H_{\text{tot}} = 0$ at the magnetization switching and $H_{\text{str}}^{\text{av}}$ is a structure field averaged over the volume of the NW under analysis. In Eq.~(\ref{Eqn:linreg}) we have introduced the scaling factor $\kappa$ to the structure field, similar to $k$ in MC simulations in Eq.~(\ref{Eqn:energy}), to take into account the effect of magnetization curling at the ends of the NWs (and different defects existing in the real sample) leading to decrease of the stray  field.

	The collected values of $H_{\text{str}}^{\text{av}}$ and related experimental $H_{\text{ext}}$ fields at selected points of the hysteresis loop for PS, QPS and various $s_y$ form a functional dependence $H_{\text{ext}} (H_{\text{str}}^{\text{av}})$.  Then, Eq.~(\ref{Eqn:linreg})  can be treated as the equation of linear regression. Thus, we can determine $\kappa$ from a slope of the line approximating the function $H_{\text{ext}}'(H_{\text{str}}^{\text{av}})$ and $H_{\text{sw}}$ from intercept of the regression line with the $H_{\text{ext}}$ axis. 
	The obtained values of $\kappa$ can be compared with the $k$ factor in the MC simulations, and $H_{\text{sw}}$ with the anisotropy field obtained from MS.

	\begin{figure}[!t]
		\centering
		\includegraphics[width=1\linewidth]{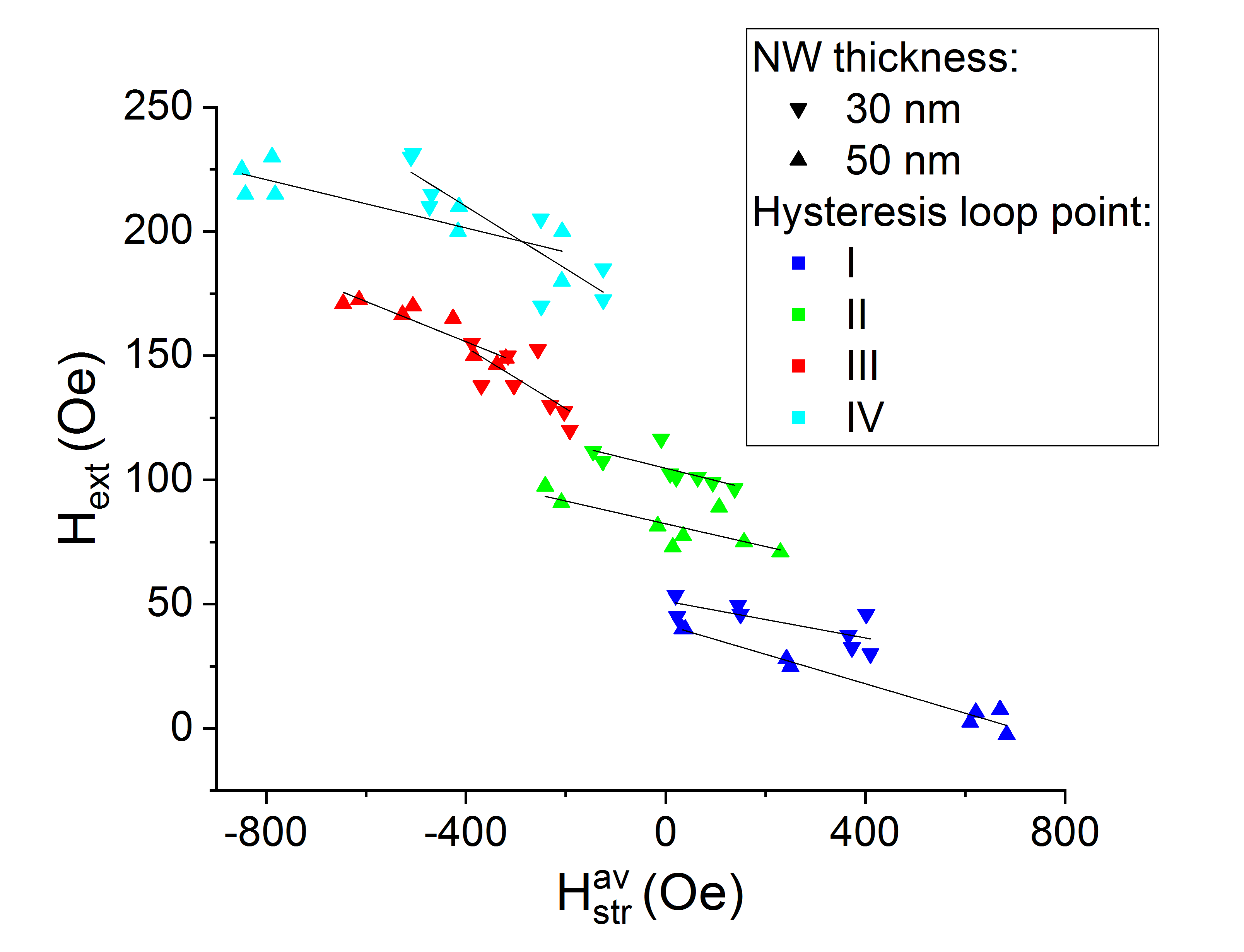}
		\caption{Dependence of the external magnetic field at the special points of the experimental hysteresis loop (I, II, III and IV) where switching of the selected NW happens on the respective structure field calculated from the analytical model. The plotted lines are calculated regression lines from which the values of $\kappa$ and $H_{\text{sw}}$ were extracted and collected in the Tab.~\ref{Tab:parameters}. The results are shown for samples investigated experimentally in the paper.}
		\label{Fig:linreg}
	\end{figure}
	
	The experimental values of $H_{\text{ext}}$ and the values of $H_{\text{str}}^{\text{av}}$ at the magnetization switching at the characteristic points of the hysteresis loop (points I to IV), and for various structures, are collected in Fig.~\ref{Fig:linreg}. The values of $\kappa$ and $H_{\text{sw}}$ obtained from linear regression analysis are collected in Tab.~\ref{Tab:parameters}. Values of the $\kappa$ parameter are compatible with $k$ extracted from MC hysteresis loops. It confirms the presence of nonuniform magnetization in NWs, which strongly reduce dipolar interactions between them. Results show that $\kappa$ does not depend strongly on dimensions of NWs (in Fig. we distinguished two thicknesses 30 and 50 nm) and the separation between the ribbons. Nevertheless, $\kappa$ has higher value in remagnetization of the narrow than of wide NWs, which points at large coupling at high magnetic fields. In most cases, $\kappa$ is higher at the beginning than at the end of the magnetization reversal process in the NWs of given width. We can point out, that stronger interactions between NWs yield faster remagnetization process. 
	
	The values of $H_{\text{sw}}$ from regression analysis are much lower in comparison to magnetic anisotropy obtained from MS. We can attribute it to the edge roughness and remagnetization process through magnetization rotation at the NW edges, which can influence anisotropy and consequently decrease $H_{\text{sw}}$.\cite{gadbois,kirk} In each case, $H_{\text{sw}}$ obtained from the regression is higher at the end than at the beginning of the magnetization reversal process in NWs of given width when it should be constant for all NWs of the same geometry in the ideal structure. This result can lead to the conclusion, that the experimental switching field can differ between NWs, which could be the effect of different level of defects in NWs.\cite{bryan,deak}

	\section{Conclusions}
	
	We have investigated experimentally the hysteresis loops for the arrays of Py NWs in dependence on the strength of magnetostatic interactions between the NWs and the type of the NW arrangement. We have studied chains with periodic and quasiperiodic sequences of wide and narrow NWs. We have molded the interactions by fabrication the arrays differing in length, thickness of the NWs, and the distance between the adjacent ribbons, i.e., the chains of the NWs. We have conducted the numerical studies based on the Monte Carlo simulations for the macrospin Ising model. The numerical computations and experimental studies have been supplemented by detailed analytical investigations of dipolar fields at different points of the hysteresis loop. 
	
	We have shown, that the interactions in the system of dipolarly coupled and ordered NWs, and thus the remagnetization process, can be controlled by various geometrical parameters. The most relevant changes have been obtained by varying the separation between the ribbons. With decreasing the separation the magnetostatic coupling between the NWs in the ribbon is significantly reduced. Moreover, the results show, that the influence of the neighboring ribbons on the remagnetization can be detectable even at separation as large as 10 $\mu$m. The change of the NW thickness and length offer the other possibilities to influence the magnetostatic interactions in the chain of NWs.
	
	We have found some differences between remagnetization processes in the periodic and Fibonacci sequences of NWs. The main difference results from the presence of the pairs of wide NWs in the quasiperiodic structure, where due to the preferential anti-parallel orientation of the magnetization in those pairs, the additional step in the hysteresis loop can exists. There are also more subtle effects demonstrated in Monte Carlo simulations, which however are hindered in experiment due to weakened magnetostatic coupling. Reduction of defects and optimization of the shape can enhance coupling and enable experimental observation of these features in hysteresis loops characteristic for QPS.

	
	\vspace{5mm}
	
	\section*{Acknowledgements}
	
	The research received funding from the Innovation Programme under Marie Sklodowska-Curie grant agreement No.~644348 (MagIC) and by the National Science Centre Poland for OPUS Grant No. 2015/17/B/ST3/00118(Metasel). J. W. K. and A. M. would like to acknowledge the financial support of the Foundation of Alfried Krupp Kolleg Greifswald and the Russian Foundation for Basic Research (research project No. 18-32-00713), respectively.
	
	\bibliographystyle{apsrev4-1}
	\bibliography{bibliography}
	
\end{document}